
\input harvmac
\noblackbox
\newcount\figno
\figno=0
\def\fig#1#2#3{
\par\begingroup\parindent=0pt\leftskip=1cm\rightskip=1cm\parindent=0pt
\baselineskip=11pt
\global\advance\figno by 1
\midinsert
\epsfxsize=#3
\centerline{\epsfbox{#2}}
\vskip 12pt
\centerline{{\bf Figure \the\figno} #1}\par
\endinsert\endgroup\par}
\def\figlabel#1{\xdef#1{\the\figno}}
\def\pano{\par\noindent}
\def\smno{\smallskip\noindent}
\def\meno{\medskip\noindent}
\def\bigno{\bigskip\noindent}
\font\cmss=cmss10
\font\cmsss=cmss10 at 7pt
\def\rlx{\relax\leavevmode}
\def\inbar{\vrule height1.5ex width.4pt depth0pt}
\def\IC{\relax\,\hbox{$\inbar\kern-.3em{\rm C}$}}
\def\IR{\relax{\rm I\kern-.18em R}}
\def\IN{\relax{\rm I\kern-.18em N}}
\def\IP{\relax{\rm I\kern-.18em P}}
\def\ZZ{\rlx\leavevmode\ifmmode\mathchoice{\hbox{\cmss Z\kern-.4em Z}}
 {\hbox{\cmss Z\kern-.4em Z}}{\lower.9pt\hbox{\cmsss Z\kern-.36em Z}}
 {\lower1.2pt\hbox{\cmsss Z\kern-.36em Z}}\else{\cmss Z\kern-.4em Z}\fi}
\def\narrowplus{\kern -.04truein + \kern -.03truein}
\def\narrowminus{- \kern -.04truein}
\def\narrowminussub{\kern -.02truein - \kern -.01truein}
\def\cl{\centerline}
\def\a{\alpha}

\def\o#1{\overline{#1}}

\def\ra{\rangle}

\def\typeo{type$\ {\rm 0B}$}

\def\typep{type$\ {\rm 0}$}
\def\typepp{type$\ {\rm 0}'$}

\def\typeb{type$\ {\rm IIB}$}



\lref\rframp{P.H. Frampton and C. Vafa, {\it Conformal Approach to Particle
   Phenomenology}, hep-th/9903226;
    P.H. Frampton, {\it Conformality and Gauge Coupling Unification},
    hep-th/9905042;
    C. Cs\' aki, W. Skiba and J. Terning, {\it $\beta$ Functions
  of Orbifold Theories and the Hierarchy Problem}, hep-th/9906057.}

\lref\rbergman{O. Bergman and M.R. Gaberdiel, {\it Dualities of Type 0
Strings}, hep-th/9906055.}

\lref\rbillo{M. Billo, B. Craps and F. Roose, {\it On D-branes in
Type 0 String Theory}, hep-th/9902196.}

\lref\rbfla{R. Blumenhagen, A. Font and D. L\"ust, {\it Tachyon-free
Orientifolds of Type 0B Strings in Various Dimensions}, hep-th/9904069.}

\lref\rbflb{R. Blumenhagen, A. Font and D. L\"ust, {\it Non-Supersymmetric 
Gauge Theories from D-Branes in Type 0 String Theory},
hep-th/9906101.}

\lref\roz{ M. Alishahiha, A. Brandhuber and Y. Oz, 
{\it Branes at Singularities in Type 0 String Theory}, 
     JHEP {\bf 9905} (1999) 024, hep-th/9903186.}

\lref\rkletsya{I.R. Klebanov and A.A. Tseytlin, {\it D-Branes and Dual Gauge 
theories in Type 0 Strings}, Nucl.Phys. {\bf B546} (1999) 155, 
hep-th/9811035.} 

\lref\rkletsyb{I.R. Klebanov and A.A. Tseytlin, {\it A Non-supersymmetric
Large N CFT from Type 0 String Theory}, JHEP {\bf 9903} (1999) 015,
hep-th/9901101.} 

\lref\rkletsyc{I.R. Klebanov and A.A. Tseytlin, {\it Asymptotic Freedom
and Infrared Behavior in the Type 0 String Approach to
Gauge Theory},  Nucl.\ Phys.\ {\bf B547} (1999) 143, hep-th/9812089.}

\lref\rminahan{J.A. Minahan, {\it Asymptotic Freedom and Confinement from
Type 0 String Theory}, JHEP {\bf 9904} (1999) 007, hep-th/9902074.}

\lref\rgarousi{M.R. Garousi, {\it String Scattering from D-Branes in Type 0
Theories}, hep-th/9901085.}

\lref\rzarembo{K. Zarembo, {\it Coleman-Weinberg Mechanism and Interaction
of D3-Branes in Type 0 String Theory}, hep-th/9901106.} 

\lref\rtsz{A.A. Tseytlin and K. Zarembo, {\it Effective Potentials in Non-
Supersymmetric $SU(N)\times SU(N)$ Gauge Theory and Interactions of Type 0
D3-Branes}, hep-th/9902095.}

\lref\rcosta{M.S. Costa, {\it Intersecting D-Branes and Black Holes in 
Type 0 String Theory}, JHEP {\bf 9904} (1999) 016, hep-th/9903128.}

\lref\rdienes{J.D. Blum and  K.R. Dienes, {\it Duality without Supersymmetry: 
The Case of the SO(16)$\times$SO(16) String}, 
Phys.Lett. {\bf B414} (1997) 260, hep-th/9707148;
{\it Strong/Weak Coupling Duality Relations for Non-Supersymmetric String 
Theories}, Nucl.Phys. {\bf B516} (1998) 83, hep-th/9707160.}

\lref\riban{L.E. Ib\'a\~nez, R. Rabadan and A.M. Uranga, 
{\it Anomalous U(1)s in Type I and TypeIIB D=4, N=1 string vacua}, 
Nucl.Phys. {\bf B542} (1999) 112, hep-th/9808139.}

\lref\rcraps{B. Craps and F. Roose, {\it NS Fivebranes in Type 0 String
Theory}, hep-th/9906179.}

\lref\rhorava{P. Ho\v rava and E. Witten, {\it Heterotic and Type I 
String Dynamics from Eleven Dimensions}, Nucl.Phys. {\bf B460} (1996) 506, 
hep-th/9510209.}

\lref\rkol{A. Armoni and B. Kol, {\it Non-Supersymmetric Large N Gauge
Theories from Type 0 Brane Configurations}, hep-th/9906081.}

\lref\rima{Y. Imamura, {\it Branes in Type 0/Type II Duality},
hep-th/9906090.}

\lref\rkim{J.Y. Kim, {\it Absorption of Dilaton s-Wave in Type 0B String
Theory}, hep-th/9906196.}

\lref\rseki{S. Seki, {\it Baryon Configurations in the UV and IR Regions of 
Type 0 String Theory}, hep-th/9906210.}

\lref\rkleb{I.R. Klebanov, {\it Tachyon Stabilization in the AdS/CFT 
Correspondence}, hep-th/9906220.}

\lref\rdixon{L. Alvarez-Gaume, P. Ginsparg, G. Moore and C. Vafa, {\it
An O(16)$\times$ O(16) Heterotic String}, 
Phys.Lett. {\bf B171} (1986) 155;
L. Dixon and J. Harvey, {\it String Theories in Ten Dimensions
Without Space-Time Supersymmetry}, Nucl.\ Phys.\ {\bf B274} (1986) 93;
N. Seiberg and E. Witten, {\it Spin Structures in String Theory}, 
Nucl.\ Phys.\ {\bf B276} (1986) 272. }

\lref\rbergab{O. Bergman and M.R. Gaberdiel, {\it A Non-Supersymmetric Open 
String Theory and S-Duality}, hep-th/9701137.}

\lref\rangel{C. Angelantonj, {\it Non-Tachyonic Open Descendants of the 
0B String Theory}, hep-th/9810214.}

\lref\rsagn{A. Sagnotti, {\it Some Properties of Open String Theories},
 hep-th/9509080;
A. Sagnotti, {\it Surprises in Open String Perturbation Theory},
hep-th/9702093.}

\lref\rsagbi{A. Sagnotti, M. Bianchi,
{\it On the Systematics of Open String Theories},
Phys.\ Lett.\ {\bf B247} (1990) 517.}

\lref\rgimpol{
E.G.\ Gimon and J.\ Polchinski, {\it Consistency Conditions
for Orientifolds and D-Manifolds}, Phys.\ Rev.\ {\bf D54} (1996) 1667,
hep-th/9601038.}

\lref\rgimjo{ E.G.\ Gimon and C.V.\ Johnson, {\it K3 Orientifolds},
Nucl.\ Phys.\ {\bf B477} (1996) 715, hep-th/9604129;
A. Dabholkar, J. Park, {\it Strings on Orientifolds},
Nucl.Phys. {\bf B477} (1996) 701, hep-th/9604178.}

\lref\rfisus{W.\ Fischler and L.\ Susskind, 
{\it Dilaton Tadpoles, String Condensates, and Scale Invariance}, 
Phys.\ Lett.\ {\bf B171} (1986) 383;  
{\it Dilaton Tadpoles, String Condensates, and Scale Invariance II}, 
Phys.\ Lett.\ {\bf B173} (1986) 262.}

\lref\rkaku{ Z. Kakushadze, G. Shiu, S.-H. H. Tye, {\it
   Type IIB Orientifolds, F-theory, Type I Strings on Orbifolds and 
    Type I - Heterotic Duality}, Nucl. Phys. {\bf B533} (1998) 25,
    hep-th/9804092;
   Z. Kakushadze, {\it Non-perturbative Orientifolds}, hep-th/9904007;
  Z. Kakushadze, {\it Type I on (Generalized) Voisin-Borcea Orbifolds and 
  Non-perturbative Orientifolds}, hep-th/9904211;
   Z. Kakushadze, {\it Non-perturbative K3 Orientifolds with NS-NS B-flux},
  hep-th/9905033. }

\lref\rdabol{ A. Dabolkar and J. Park, {\it An Orientifold of Type 
        IIB theory on K3}, Nucl. Phys. {\bf B472} (1996) 207, 
        hep-th/9602030;
        {\it Strings on Orientifolds}, Nucl. Phys. {\bf B477}
        (1996) 701, hep-th/9604178.}

\lref\rjohnson{E. Gimon and C. Johnson, {\it K3 Orientifolds},
        Nucl. Phys. {\bf B477} (1996) 715, 
        hep-th/9604129.}

\lref\rlowe{ S. Chaudhuri and D. Lowe, {\it Type IIA-Heterotic 
Duals with Maximal Supersymmetry}, Nucl. Phys. {\bf B459} (1996) 113,
hep-th/9508144.}

\Title{\vbox{\hbox{hep--th/9906234}
 \hbox{HUB--EP--99/29}\hbox{CERN-TH/99-190}}}
{\vbox{\centerline{A Note on Orientifolds and Dualities}
\bigskip\centerline{of Type 0B String Theory}
}}
\centerline{Ralph Blumenhagen${}^1$ and Alok Kumar${}^{2,3}$}
\bigskip
\centerline{\it ${}^{1}$ Humboldt-Universit\"at Berlin, Institut f\"ur 
Physik,}
\centerline{\it  Invalidenstrasse 110, 10115 Berlin, Germany }
\smallskip
\centerline{\it ${}^2$ Theory Divison, CERN, CH-1211}
\centerline{Geneva 23, Switzerland}
\smallskip
\centerline{\it ${}^3$ Institute of Physics,} 
\centerline{Bhubaneswar 751 005, India}
\bigskip
\bigskip
\centerline{\bf Abstract}
\noindent
We generalize the construction of four dimensional non-tachyonic 
orientifolds of 
type 0B string theory to non-supersymmetric backgrounds.
We construct a four dimensional model containing  self-dual D3 and
D9-branes and leading to a chiral anomaly-free massless spectrum. 
Moreover, we discuss a further tachyon-free six dimensional model with
only D5 branes. Eventually, we  speculate about strong coupling dual models
of the ten-dimensional orientifolds of type 0B.  
\footnote{}
{\pano
${}^1$ e--mail:\ blumenha@physik.hu-berlin.de
\pano
${}^2$ e--mail:\ Alok.Kumar@cern.ch
\pano}
\Date{06/99}

\newsec{Introduction}

Non-supersymmetric string theories have received a lot of attention during
the last year 
\refs{\rkletsya\rkletsyb\rkletsyc\rminahan\rgarousi
\rzarembo\rbillo\rtsz\rcosta\roz\rbfla\rbergman\rkol\rima\rbflb\rkim
\rseki{--}\rkleb}. 
Most of the effort went into
a better understanding of phenomena in the effective 
non-supersymmetric gauge theories
including a new approach to solve the hierarchy problem by a 
non-supersymmetric conformal gauge theory in some intermediate energy
regime \rframp.
Parallel to that there were some attempts to construct consistent 
non-supersymmetric
four dimensional string vacua. Besides the seven heterotic non-supersymmetric
ten-dimensional string theories   
there also exist the two non-supersymmetric 
type 0A/B string theories \rdixon. The latter models both contain a tachyonic 
mode which renders the theory unstable. One way to circumvent this problem is
by placing enough RR flux in the background which shifts the square of
the tachyon mass to positive values. This approach was used in 
\rkletsya\ to reliably study the effective theory on the D3 branes. 
Another way of getting rid of the tachyon is by doing a special orientifold 
\refs{\rsagn,\rsagbi,\rbergab}. 

There exist three independent orientifold projections of \typeo\ string
theory $\{\Omega,\Omega (-1)^{F_R},\Omega (-1)^{f_R} \}$, where $(-1)^{F_R}$
is the right moving space-time fermion number and $(-1)^{f_R}$ is
the right moving world-sheet fermion number. The first two models 
still contain a bulk tachyon, whereas in the third, the \typepp,  model 
the tachyon
is projected out. The massless spectrum of the resulting ten-dimensional
non-supersymmetric string theory contains the graviton, the dilaton,
a RR 0-form, a 2-form and a self-dual 4-form in the closed string sector
and a gauge field in the adjoint of U(32)  equipped by
a left-handed Majorana-Weyl
fermion in the ${\bf 496}\oplus\o{\bf 496}$ representation in the open 
string sector. The latter fermions necessarily appear for this class
of orientifolds, because the world-sheet 
fermion number
operator $(-1)^{f_R}$ exchanges a D9 brane charged under the first of
the two RR 10-forms in \typeo\ with a D9' brane charged under the second  of
the two RR 10-forms. Open strings stretched between these different kinds
of 9-branes lead to space-time fermions.

As was pointed out in \refs{\rangel,\rbfla} requiring
the absence of tachyons also for compactified \typepp\ orientifolds is highly
restrictive and so far only one $\ZZ_2$ orientifold 
in six-dimensions and one $\ZZ_3$ orientifold in four dimensions have
been constructed.
In all other cases there appear extra tachyons in twisted sectors which
can not all be projected out as long as $\Omega$ exchanges a $g$-twisted 
sector with a $g^{-1}$ twisted sector. 
One can allow  $\Omega$ to act  without exchanging twisted sectors but as
was argued in \rkaku\ such models contain extra non-perturbative states.   

All compact models studied so far were supersymmetric backgrounds in the
sense that when used in a \typeb\ compactification they preserve
some supersymmetry. However, since \typeo\ is non-supersymmetric anyway, one
might try to consider backgrounds not preserving any supersymmetry at all. 
In this case the absence of tachyons is highly restrictive, as well, but
in the first part of this letter in some detail we 
will  present one specific $\ZZ_2$ model  in which everything works
out nicely. It turns out that in order to cancel all RR tadpoles one has to
introduce D9/D9' and D3/D3' branes. Of course, as in all \typepp\
orientifolds there remains an uncancelled dilaton tadpole which can 
be cured by the Fischler/Susskind mechanism \rfisus. 
In the second part of this paper we will discuss  one more compact model 
in six dimensions followed by some speculations about possible dual
models of the ten-dimensional \typeo\ orientifolds.

\newsec{Type 0' orientifold on a non-supersymmetric background }

We take \typeo\ string theory, compactify it on a six torus $T^6=(S^1)^6$ and 
divide out by the orientifold group $\{ (1+R) + \Omega'(1+R)\}$
with $\Omega'=\Omega (-1)^{f_R}$ and $R:z_i\to -z_i$ for all $i\in\{1,2,3\}$.
Note, that in a \typeb\ compactification $R$ would not satisfy
level-matching in the NS-R and R-NS sector. However, precisely these
two sectors are absent in \typeo\ string theory 
so that dividing out by $R$ leads to
a modular invariant torus amplitude. 
Already at this stage we would like to point out that there exists a 
subtlety in the Ramond sector which will become
important in the open string sector. Since the action of $R$ on the 
left-moving
Ramond sector ground states is given by 
\eqn\chapaa{ R|s_1\, s_2\, s_3\, s_4\ra=e^{\pi i(s_2+s_3+s_4)}
                  |s_1\, s_2\, s_3\, s_4\ra =\pm i|s_1\, s_2\, s_3\, s_4\ra
                 \quad\quad {\rm with}\ 
                  s_i=\pm{1\over 2} ,}
it acts rather like a $\ZZ_4$ than a $\ZZ_2$ operation. 
Of course in the closed string sector the left-moving
Ramond sector is always paired with the right-moving
Ramond sector, so that here $R$ really acts  like a $\ZZ_2$.
As usual in orientifolds one has to compute all one-loop diagrams 
and require tadpole cancellation. 

\subsec{ The Klein bottle amplitude}

The computation of the Klein bottle amplitude is straightforward
\eqn\chapab{\eqalign{ K&= 4\, c \, \int_0^\infty {dt\over t^3}\, 
                  {\rm Tr} \left[
                  {\Omega'\over 2}\, 
            {1+R\over 2}{1+(-1)^{f_L+f_R} \over 2}\, 
                   e^{-2\pi t(L_0+\o{L}_0)} \right] 
                  \cr
&= -  c \,\int_0^\infty {dt\over t^3}\, {f_4^8(e^{-2\pi t})
                      \over f_1^8(e^{-2\pi t}) } \left(
               \left[ \sum_m e^{-\pi t {m^2\over \rho}} \right]^6 +
               \left[ \sum_n e^{-\pi t {n^2 \rho}} \right]^6 \right), \cr }}
with $c = V_{4}/(8\pi^2 \a')^2$ and $\rho=r^2/\alpha'$. 
The transformation to tree channel 
reveals  that there are RR 10-form and RR 4-form tadpoles but no NS tadpole.

\subsec{ The annulus amplitude}

In order to cancel these RR tadpoles we introduce $N_9$ pairs of D9/D9'
branes and $N_3$ pairs of D3/D3' branes. 
Note, that in a type I  compactification this would
be impossible,  as in contrast to \typepp\ string theory 
type I string theory simply  does not contain any D3-branes. 
Since open strings stretched between a Dp and Dp' brane have
fermionic zero modes, we are facing the subtlety mentioned in \chapaa.
The action of $R$ on the modes leads to extra factors of $i$ which must
be compensated by further factors of $i$ from the action of $R$ on the
Chan-Paton factors of the open strings. 
Thus, in the open string sector we better consider $R$ as a $\ZZ_4$ action,
so that the annulus amplitude is 
\eqn\chapac{ A= c \, \int_0^\infty {dt\over t^3}\, 
                  {\rm Tr} \left[
                  {1\over 2}\ 
            {1+R+R^2+R^3\over 4}\ {1+(-1)^{F_s} \over 2}\ {1+(-1)^{f} \over 2}
             \, e^{-2\pi t L_0} \right] }
where the trace has to be taken over all open strings stretched between
the four different kinds of D-branes $\{D9,D9',D3,D3'\}$. 
The presentation of the whole amplitude would be much too lengthy to present
here. However, completely analogous to the model discussed in \rgimpol, 
cancellation of the  twisted RR tadpoles requires
\eqn\chapad{  {\rm Tr}(\gamma_{R,p})=0 \quad\quad {\rm with}\ 
                p\in\{9,9',3,3'\}  .}
Neglecting all the terms becoming zero by the choice in \chapad,
for instance  in the
99, 9'9', 99' and 9'9 sectors the total annulus amplitude reads 
\eqn\chapae{\eqalign{ A_{99}=c\int {dt\over t^3} 
        &\left[ \sum_m e^{-2\pi t {m^2\over \rho}} \right]^6\
             \biggl[
               {N_9^2\over 8} \left(f_3^8-f_4^8-f_2^8\over f_1^8\right)
                     (e^{-\pi t}) +\cr
                 &{1\over 16}\left[ {\rm Tr}(\gamma_{R^2,9})                  
                   {\rm Tr}(\gamma^{-1}_{R^2,9}) + 
                   {\rm Tr}(\gamma_{R^2,9'})                  
                   {\rm Tr}(\gamma^{-1}_{R^2,9'})\right]
                  \left(f_3^8-f_4^8 \over f_1^8\right)(e^{-\pi t}) +\cr 
                 &{1\over 8} \  {\rm Tr}(\gamma_{R^2,9})                  
                   {\rm Tr}(\gamma^{-1}_{R^2,9'})\ 
                   {f_2^8 \over f_1^8}(e^{-\pi t}) \biggr] .\cr }}
For open strings stretched between D3/D3' branes the result is completely
analogous.  
If we would simply choose 
${\rm Tr}(\gamma_{R^2,9})={\rm Tr}(\gamma_{R^2,9'})=0$ then only the first 
term in \chapae\ would be non-zero. However, since we have introduced
in \chapac\ an extra factor of two in the denominator compared to a usual
$\ZZ_2$ annulus amplitude, we would get a problem with tadpole cancellation. 
We can cure this missing factor of two by requiring that the second and third
term in \chapae\ add up exactly to the first term
leading to 
\eqn\chapaf{\eqalign{  &{\rm Tr}(\gamma_{R^2,9})                  
                   {\rm Tr}(\gamma^{-1}_{R^2,9}) = 
                   {\rm Tr}(\gamma_{R^2,9'})                  
                   {\rm Tr}(\gamma^{-1}_{R^2,9'})=N_9^2 \cr
                 & {\rm Tr}(\gamma_{R^2,9})                  
                   {\rm Tr}(\gamma^{-1}_{R^2,9'})=-N_9^2 .\cr}}
A solution to the conditions \chapad\ and \chapaf\ is 
\eqn\chapag{  \Gamma_{R,9}= \left(\matrix{ \gamma_{R,9} & 0 \cr
                           0 & \gamma_{R,9'} \cr}\right)_{2N_9,2N_9}=
                        \left(\matrix{  I  &  0  & 0 & 0 \cr
                                        0  & -I & 0 & 0 \cr
                                        0  & 0  &  i\, I  &  0  \cr
                                        0 & 0 &  0 & -i\, I    \cr
                   } \right)_{2N_9,2N_9} }
where $I$ denotes the matrix $I={\rm diag}[1,\ldots,1]$.
With this choice of Chan-Paton actions the non-zero contribution of the open 
strings stretched between the various 9 and 3 branes to the annulus
amplitude is
\eqn\chapah{ A_{93}=c\int {dt\over t^3} {1\over 2}
               N_9 N_3 \left( {f_3^2 f_2^6 - f_2^2 f_3^6 \over f_1^2 f_4^6}
                \right)(e^{-\pi t})  .}
As we will see below in tree channel this leads to further NSNS tadpoles.

\subsec{ The M\"obius amplitude}

In order to compute the M\"obius amplitude
\eqn\chapai{ M= c \, \int_0^\infty {dt\over t^3}\, 
                  {\rm Tr}_{99',33'} \left[
                  {\Omega'\over 2}\ 
            {1+R+R^2+R^3\over 4}\ {1+(-1)^{F_s} \over 2}\ {1+(-1)^{f} \over 2}
             \, e^{-2\pi t L_0} \right] }
we have to take into account  that $\Omega'$ acts on the ground states in 
the 99' and 33' sector in the following way
\eqn\chapaj{\eqalign{ &\Omega' |s_1\, s_2\, s_3\, s_4\ra_{99'}=
          -|s_1\, s_2\, s_3\, s_4\ra_{99'} \cr
          &\Omega' |s_1\, s_2\, s_3\, s_4\ra_{33'}=
                -e^{\pi i(s_2+s_3+s_4)}
                  |s_1\, s_2\, s_3\, s_4\ra_{33'} .\cr}}
For the 99' sector there are non-zero contributions from the $\Omega'$ and
$\Omega'R^2$ insertions in the trace. With the choice of the
$\Gamma_R$ matrix above one gets
\eqn\chapak{ M_{99'}=c \, \int_0^\infty {dt\over t^3}\, {1\over 8}
                    {\rm Tr}(\Gamma_{\Omega',9}^T\, \Gamma_{\Omega',9}^{-1})
                   \ {f_2^8(i\, e^{-\pi t})\over f_1^8(ie^{-\pi t})}
                 \left[ \sum_m e^{-2\pi t {m^2\over \rho}} \right]^6. }
For the 33' sector there are non-zero contributions from the $\Omega'R$ and
$\Omega'R^3$ insertions in the trace. With the choice of the
$\Gamma_R$ matrix above one gets
\eqn\chapak{ M_{33'}=c \, \int_0^\infty {dt\over t^3}\, {1\over 8}
                    {\rm Tr}(\Gamma_{\Omega'R,3}^T\, \Gamma_{\Omega'R,3}^{-1})
                   \ {f_2^8(i\, e^{-\pi t})\over f_1^8(ie^{-\pi t})}
                 \left[ \sum_n e^{-2\pi t {n^2 \rho}} \right]^6. }
Thus, the M\"obius amplitude  only leads  to RR tadpoles in the tree channel.

\subsec{Tadpole Cancellation}

Transforming all the amplitude in tree channel and extracting the divergent
pieces one derives  the following two RR tadpole cancellation conditions 
\eqn\chapal{\eqalign{   V_4 \rho^3 &\left( N_9^2-
              32\, {\rm Tr}(\Gamma_{\Omega',9}^T\, \Gamma_{\Omega',9}^{-1}) 
              +2^{10} \right)=0 \cr
               V_4/\rho^3 &\left( N_3^2-
              32\, {\rm Tr}(\Gamma_{\Omega'R,3}^T\, \Gamma_{\Omega'R,3}^{-1}) 
              +2^{10}\right)=0 .\cr }}
The first equation tells us that $\Gamma_{\Omega',9}$ is symmetric, where
we can always make the choice
\eqn\chapam{  \Gamma_{\Omega',9}= 
                        \left(\matrix{  0  &  0  & I & 0\cr
                                        0  & 0 & 0 & I\cr
                                        I & 0 &  0  &  0  \cr
                                        0 & I &  0 & 0   \cr
                   } \right)_{2N_9,2N_9} }
so that $N_9=32$. T-duality as well as the algebra of $\gamma$ matrices 
leads to the following choice of
$\Gamma_{\Omega',3}$ and $\Gamma_{R,3}$ 
\eqn\chapam{  \Gamma_{\Omega',3}= 
                        \left(\matrix{  0  &  0  & -i\, I & 0\cr
                                        0  & 0 & 0 & i\, I\cr
                                        I & 0 &  0  &  0  \cr
                                        0 & -I &  0 & 0   \cr
                   } \right)_{2N_3,2N_3},\quad\quad
              \Gamma_{R,3}= 
                        \left(\matrix{  i\, I  &  0  & 0 & 0\cr
                                        0  & -i\, I & 0 & 0\cr
                                        0 & 0 &   I  &  0  \cr
                                        0 & 0  &  0 &  -I  \cr
                   } \right)_{2N_3,2N_3}, }
which indeed satisfies
\eqn\chapan{ \Gamma_{R,3}\Gamma_{\Omega',3}\sim\Gamma_{\Omega',3}
                     (\Gamma_{R,3}^{-1})^T\sim\Gamma_{\Omega',9} }
up to phases implying that $N_3=32$.
As always in \typepp\ orientifolds there remains an uncancelled NSNS 
tadpole 
\eqn\chapao{ 8c \int_0^\infty dl\  V_4\, \left( N_9^2 \rho^3 + N_3^2/\rho^3 -
                       N_9 N_3/8 \right) .}
which needs to be cancelled by the Fischler/Susskind mechanism \rfisus.

\subsec{The massless spectrum}
 
Having defined all the actions of the $R$ and $\Omega'$ on the various
modes and on the Chan-Paton factors it is now a straightforward
exercise to compute the massless spectrum as displayed 
for the closed string sector in Table 1.
\meno
\cl{\vbox{
\hbox{\vbox{\offinterlineskip
\def\tablespace{height2pt&\omit&&
 \omit&\cr}
\def\tablerule{\tablespace\noalign{\hrule}\tablespace}

\hrule\halign{&\vrule#&\strut\hskip0.2cm\hfil#\hfill\hskip0.2cm\cr
\tablespace
& sector && field      &\cr
\tablerule
& untwisted NS-NS && $G_{\mu\nu}$, dilaton $\Phi$, 21 scalars &\cr
\tablespace
& untwisted R-R && 32 scalars  &\cr
\tablerule
& twisted NS-NS && massive &\cr
\tablespace
& twisted R-R && 64 scalars &\cr
\tablespace}\hrule}}}}
\cl{
\hbox{{\bf Table 1:}{\it ~~ Closed string spectrum of $T^6/\ZZ_2$ }}}
\meno
Thus, there is the graviton, the dilaton and 117 further scalars.
For the case that all 32 D3/D3' branes are placed on the same fixed point
of $R$ on $T^6$, we derive the massless open string spectrum listed in 
Table 2.
\meno
\cl{\vbox{
\hbox{\vbox{\offinterlineskip
\def\tablespace{height2pt&\omit&&\omit&&
 \omit&\cr}
\def\tablerule{\tablespace\noalign{\hrule}\tablespace}

\hrule\halign{&\vrule#&\strut\hskip0.2cm\hfil#\hfill\hskip0.2cm\cr
\tablespace
& sector && field && gauge    U(16)$\times$ U(16)$\vert_{9}\times$ U(16)
$\times$  U(16)$\vert_3$  &\cr
\tablerule
& 99, 33  && vector && {\bf Adj} &\cr
\tablespace
& $9'9'$, $3'3'$ && scalar && $6\times\{ ({\bf 16},{\bf\o{16}};
    {\bf 1},{\bf 1}) + 
   ({\bf\o{16}}, {\bf{16}};{\bf 1},{\bf 1}) + 
   ({\bf 1},{\bf 1};{\bf 16},{\bf\o{16}}) + 
 ({\bf 1},{\bf 1};{\bf\o{16}},{\bf 16}) \}$ &\cr 
\tablerule
& 93, $9'3'$  && -- &&  massive &\cr
\tablerule
& $99'$, $33'$ && Weyl && 
  $4\times\{ ({\bf 120} ,{\bf 1};{\bf 1},{\bf 1}) + 
        ({\bf 1},{\bf 120};{\bf 1},{\bf 1}) + 
  ({\bf 1},{\bf 1};{\bf 120},{\bf 1})+
    ({\bf 1},{\bf 1};{\bf 1},{\bf 120}) \}$ &\cr
\tablespace
& && Weyl && $4\times\{ ({\bf\o{16}},{\bf\o{16}};{\bf 1},{\bf 1}) + 
     ({\bf 1},{\bf 1};{\bf\o{16}},{\bf\o{16}}) \}$ &\cr 
\tablerule
& $93'$, $39'$ && Weyl && $\{ ({\bf 16},{\bf 1};{\bf 16},1) + 
               ({\bf 1},{\bf 16};{\bf 1},{\bf 16}) \} $ &\cr 
\tablespace}\hrule}}}}
\cl{
\hbox{{\bf Table 2:}{\it ~~ Open string spectrum of $T^6/\ZZ_2$ }}}
\meno 
Note, that the  chiral  massless spectrum in Table 2
passes the non-trivial check of absence of non-abelian gauge 
anomalies. Similar to the supersymmetric case we expect 
the anomalous U(1)s to be cancelled by some generalized Green-Schwarz
mechanism \riban.

\newsec{Type 0' orientifold in 6D with D5-branes}

We now discuss the possibility of constructing 
type 0' orientifolds using certain symmetry properties of the  
compactified manifolds. In particular, 
orientifolds of type IIB on K3 have been constructed 
previously \refs{\rdabol,\rjohnson} 
and give rise to anomaly free models in six dimensions
with different number of tensor, vector and hyper-multiplets. 
It is also known that K3 possesses certain discrete isometries
which can be combined with orientifold projections. Since
these isometries maintain supersymmetry, one is able to obtain 
consistent models without breaking supersymmetry further. Such 
projections have also been utilized earlier for constructing 
type II examples of string models which are dual to the 
reduced rank heterotic string compactifications with maximal 
supersymmetry \rlowe. 

We now show that such projections can also be applied to construct
consistent type 0' orientifolds. As an explicit example we concentrate on a
particular $\ZZ_2$ isometry of K3 which leaves all three self-dual 
2-forms of K3 invariant. However among the 19 anti-self-dual 
2-forms 8 are odd under its operation. We work with a particular
realization of this projection in the orbifold limit ${T^4/\ZZ_2}$ of 
K3 \rdabol. The orbifold is constructed
by complex coordinates $(z_1, z_2)$ on the torus $T^4$ defined 
by periodic identifications: $z_1 \sim z_1 + 1$, 
$z_1 \sim  z_1 + i$ (similarly for $z_2$) and by dividing the
torus by $\ZZ_2 \equiv \{1, R\}$, where the projection $R$
acts on complex coordinates as, $R:$ 
$(z_1, z_2) \rightarrow (-z_1, -z_2)$. On the other hand the 
isometry of K3 mentioned above is represented by an operation
$S:$ $(z_1, z_2) \rightarrow (-z_1 + {1\over 2}, -z_2 + {1\over 2})$. 

The orientifold model that we are considering involves 
the projections $R$ and $\Omega' S$. The closed string spectrum in 
the untwisted NS-NS sector then consists of the graviton, dilaton 
and 10 scalars denoted by the representations 
$[(3, 3) + 11 (1, 1)]$ of 
$SU(2)\times SU(2)$, which is the little group of the Lorentz group
in six dimensions. The untwisted R-R sector consists of 4 self-dual 
and 4 anti-self-dual 2-forms, in addition to 8 scalars, 
represented as: $[ 4 (3, 1) + 4 (1, 3) + 8 (1, 1)]$. 
The twisted NS-NS sectors contribute 64 scalars and the twisted R-R
sectors contribute states: $[ 8 (3, 1) + 8 (1, 3) + 16 (1, 1)]$. 

As a result the total closed string spectrum consists of 
graviton, dilaton, 98 scalars and 12 self-dual as well as 12
anti-self-dual 2-forms. The model is free of gravitational anomaly,
as the self-dual and anti-self-dual 2-forms come in equal numbers.

To obtain the open string spectrum, one analyses the tadpole 
cancellation conditions. Twisted RR tadpoles are once again 
cancelled by choosing $\gamma_R$ traceless for both 9 and 
5-branes. Since the action of S includes 
shifts along coordinates $x_6$ and $x_8$, the model is also free of
the massless RR 10-form tadpoles in the untwisted sector. 
As a result, apart from uncancelled NS-NS tadpoles, one
only has RR 6-form tadpoles which are cancelled by adding only 32 
D5 and 32 D5'-branes in two possible ways:

\item{(i)} 16 5-branes and 16 5'-branes at a fixed point $y$ of $R$ and 
the same numbers of them at the image of $y$ under $S$. The Chan-Paton
indices are determined by a $32\times 32$ block-diagonal matrix 
$\gamma_R = {\rm diag}[M, M]$ with $M$ a $16\times 16$ matrix:
$M = {\rm diag}[I_8, -I_8]$. The resulting open string spectrum 
is listed in Table 3.
\vskip 0.3cm
\cl{\vbox{
\hbox{\vbox{\offinterlineskip
\def\tablespace{height2pt&\omit&&\omit&&
 \omit&\cr}
\def\tablerule{\tablespace\noalign{\hrule}\tablespace}

\hrule\halign{&\vrule#&\strut\hskip0.2cm\hfil#\hfill\hskip0.2cm\cr
\tablespace
& sector && spin && gauge    U(8)$\times$ U(8)$\times$ U(8)'
$\times$  U(8)'  &\cr
\tablerule
& 55, 5'5'  && vector && {\bf Adj} &\cr
\tablespace
&   && $(1,1)$ && $4\times\{ ({\bf 8},{\bf\o{8}};
    {\bf 1},{\bf 1}) + 
   ({\bf\o{8}}, {\bf{8}};{\bf 1},{\bf 1}) + 
   ({\bf 1},{\bf 1};{\bf 8},{\bf\o{8}}) + 
 ({\bf 1},{\bf 1};{\bf\o{8}},{\bf 8}) \}$ &\cr 
\tablerule
& $55'$  && $(1,2)$ && 
        $2\times\{ ({\bf 8},{\bf 1};{\bf 8},{\bf 1}) + 
             ({\bf\o{8}},{\bf 1};{\bf\o{8}},{\bf 1}) 
             + ({\bf 1},{\bf 8};{\bf 1},{\bf 8})+
               ({\bf 1},{\bf\o{8}};{\bf 1},{\bf\o{8}}) \} $ &\cr
& && $(2,1)$ && 
       $2\times\{ ({\bf 8},{\bf 1};{\bf 1},{\bf 8}) + 
            ({\bf\o{8}},{\bf 1};{\bf 1},{\bf\o{8}}) 
             + ({\bf 1},{\bf 8};{\bf 8},{\bf 1})+
               ({\bf 1},{\bf\o{8}};{\bf\o{8}},{\bf 1}) \} $ &\cr
\tablespace}\hrule}}}}
\cl{
\hbox{{\bf Table 3:}{\it ~~ Open string spectrum of $T^4/\ZZ_2$ }}}
\smno
\item{\phantom{(i)}}
Since there are equal number of left and right-handed 
fermions, the model is anomaly free. \smno

\item{(ii)} Alternatively, one can take 16 5-branes (as well as 16 5'-branes)
to lie at a fixed point $x$ of $S$ and the remaining ones at the image
of $x$ under $R$. In this case the gauge group is U(16). 
To compare, for a similar IIB model the gauge group was 
$SO(16)$ \refs{\rdabol} and the difference comes due to the 
form of $\gamma_{\Omega' S}$. In our case $\gamma_{\Omega' S}$ is a 
$32\times 32$ matrix with identities ($I_{16}$) along the off-diagonal 
blocks, so as to mix the 5 and 5'-branes under $\Omega'$. 
In this case, one obtains the massless spectrum shown in Table 4.
\vskip 0.3cm
\cl{\vbox{
\hbox{\vbox{\offinterlineskip
\def\tablespace{height2pt&\omit&&\omit&&
 \omit&\cr}
\def\tablerule{\tablespace\noalign{\hrule}\tablespace}

\hrule\halign{&\vrule#&\strut\hskip0.2cm\hfil#\hfill\hskip0.2cm\cr
\tablespace
& sector && spin && gauge    U(16)  &\cr
\tablerule
& 55, 5'5'  && vector && {\bf Adj} &\cr
\tablespace
&   && $(1,1)$ && $4\times  {\bf Adj} $ &\cr 
\tablerule
& $55'$  && $(1,2)$ && 
        $2\times\{ {\bf 120}+{\bf\o{120}} \}$ &\cr
& && $(2,1)$ && 
       $2\times\{ {\bf 120}+{\bf\o{120}} \}$ &\cr
\tablespace}\hrule}}}}
\cl{
\hbox{{\bf Table 4:}{\it ~~ Open string spectrum of $T^4/\ZZ_2$ }}}
The model is once again anomaly free. 
\smno
In this section we have presented one tachyon free six dimensional 
type 0' orientifold using a discrete isometry of  
K3. As mentioned earlier, K3 has several other discrete
isometries, knows as finite abelian automorphism groups. One  can 
use these isometries as projection elements to construct other models
as well. One can also use certain supersymmetry non-preserving
involution of K3, known as Enrique involution, to construct 
a type 0' model. However it turns out that the Klein-Bottle 
amplitude is free of tadpoles and hence there is no need to 
add any D-branes. The model is found  to be 
purely bosonic in  the closed string sector. 

\newsec{A duality conjecture}

In \rbergman\ a very intriguing conjecture about the strong coupling limit
of type 0A was presented, 
namely that type 0A string theory at strong coupling is
M-theory compactified on $S^1/(-1)^{F_s}S$, where $S$ denotes the half-shift
along the compact circle. So far no dual candidate has been conjectured
for the \typepp\ orientifold. In \rbergab\ arguments were given
that the \typep\ orientifold with gauge group SO(32)$\times $SO(32)
might be dual to the bosonic string compactified
on an SO(32) torus. In the following we would like to propose
a dual model for the \typep\ orientifold with 
gauge group (SO(16)$\times $SO(16)$)^2$, which also contains space-time
fermions\footnote{$^1$}{We thank M. Gaberdiel for some comments about his
point.}.
Analogous to the discussion in \rhorava, consider M-theory 
compactified to nine dimensions on $S^1/\ZZ_2\times S^1/(-1)^{F_s}S$, 
where the first $\ZZ_2$ is the reflection $I_1:x_{10}\to -x_{10}$ 
combined with the change of the sign of the three-form. 
Compactifying first $x_{10}$ and then $x_9$ one arrives at the
$E_8\times E_8$ heterotic string compactified on $S^1/(-1)^{F_s}S$.
As argued in \rdienes, after applying T-duality for $r_{9}\to 0$ 
this should be the ten-dimensional non-tachyonic SO(16)$\times$ SO(16)
heterotic string. Exchanging the role of $x_{10}$ and $x_9$
one arrives at type 0A/$\Omega\, I_1$, which is supposed to be T-dual to
type 0B/$\Omega$. 
Taking this chain  of dualities  seriously, one is led to the following
strong-weak duality conjecture:
\meno
\cl{ {\bf Type 0 orientifold}
     $\buildrel {\rm ``dual''}\over \Longleftrightarrow$  
    {\bf non-susy SO(16)$\times$ SO(16) heterotic string.} }
\meno
At first glance this seems to make no sense, as the gauge group of 
the \typep\ orientifold has rank equal to 32. However, as was noticed
in \refs{\rbergman,\rcraps}, at strong
coupling the pair of a Dp and a Dp' brane might form a bound state
in which half of the zero modes at weak
coupling are frozen. By T-duality this implies that also the rank
of the gauge group get reduced by a factor of two at strong coupling. 
Therefore, we think the conjecture not necessarily is nonsence 
and it deserves a more detailed investigation about the actual sense 
in which the duality relation should be understood. 
In view of the duality above, it is very tempting to make
the following conjecture for the \typepp\ orientifold:
\meno
\cl{ {\bf Type 0' orientifold} 
     $\buildrel {\rm ``dual''}\over \Longleftrightarrow$  
     {\bf non-susy U(16) heterotic string.} }
\meno
The U(16) heterotic model has a tachyon which makes the whole
picture somehow symmetric. 
In the former dual pair the heterotic model encounters a tachyonic 
instability at strong coupling whereas in the latter dual pair it is 
the orientifold model which has a tachyonic instability at strong coupling.
The analysis of interpolating models carried out in \rdienes\ suggests
that all these models are dynamically driven to their supersymmetric 
counterparts. 
\centerline{{\bf Acknowledgements}}\pano
R.B. thanks Anamaria Font and Dieter L\"ust for encouraging discussions.
A.K. thanks the CERN Theory division for hospitality. 
\bigno

\listrefs
\bye